\documentclass[aps,prb,twocolumn,showpacs,preprintnumbers,amsmath,amssymb]{revtex4}
\usepackage{graphicx}

\begin{document}

\title{The Origin of Anomalous Low-Temperature Downturns in the Thermal Conductivity of Cuprates}
\author{M. F. Smith$^{\dagger}$} \author{Johnpierre Paglione$^{\dagger\dagger}$} \author{M. B.
Walker$^{\ddagger}$} \affiliation{Department of Physics, University of Toronto, Toronto, Canada.}
\author{Louis Taillefer$^{\ddagger}$}\affiliation{D\'epartement de physique et Regroupement qu\'eb\'ecois sur les mat\'eriaux de pointe, Universit\'e de Sherbrooke, Sherbrooke, Canada}

\date{\today}

\begin{abstract} We show that the anomalous decrease in the thermal
conductivity of cuprates below 300 mK, as has been observed recently in several cuprate materials
including Pr$_{2-x}$Ce$_x$CuO$_{7-\delta}$ in the field-induced normal state, is due to the thermal
decoupling of phonons and electrons in the sample.  Upon lowering the temperature, the
phonon-electron heat transfer rate decreases and, as a result, a heat current bottleneck develops
between the phonons, which can in some cases be primarily responsible for heating the sample, and the
electrons.  The contribution that the electrons make to the total low-$T$ heat current is thus
limited by the phonon-electron heat transfer rate, and falls rapidly with decreasing temperature,
resulting in the apparent low-$T$ downturn of the thermal conductivity.  We obtain the temperature
and magnetic field dependence of the low-$T$ thermal conductivity in the presence of phonon-electron
thermal decoupling and find good agreement with the data in both the normal and superconducting
states.

\end{abstract}


\maketitle

\section{Introduction}

In recent measurements made of the thermal conductivity $\kappa$ of high $T_C$ cuprates at
milliKelvin temperatures, an unexplained decrease in the electronic component of thermal conductivity
with decreasing temperature well below 1 K has been observed (we will refer to such decreases as
``downturns''). The most dramatic observation of a downturn in $\kappa/T$ occurred in the field
induced normal state of optimally doped Pr$_{2-x}$Ce$_x$CuO$_{7-\delta}$ (henceforth PCCO), where the
electronic contribution $\kappa_{el}/T$ was found to decrease below 300 mK and vanish at the lowest
measured temperatures\protect\cite{rhill}. This was in startling contrast to electrical conductivity
measurements made on the same material (and to NMR studies \protect\cite{zheng}), which showed
apparently normal metallic behaviour over the same temperature range.  These measurements appeared to
show an unexpected violation of the Wiedemann-Franz law at very low temperatures\protect\cite{rhill},
with heat conductivity becoming much less than charge conductivity.  (Note that the main effect,
observed above 300 mK, is a violation whereby heat conductivity is greater than charge conductivity
by a factor of 1.7-1.8.)\protect\cite{rhill}

Downturns have been observed in other cuprate materials in both the normal and superconducting
states.  For example, data on non-superconducting, overdoped La$_{2-x}$Sr$_x$CuO$_4$ and on optimally
and overdoped samples of superconducting La$_{2-x}$Sr$_{x}$CuO$_4$  and YBa$_2$Cu$_3$O$_{7-\delta}$
(henceforth LSCO and YBCO, respectively) have showed similar downturns \protect\cite{nakamae}
\protect\cite{takeya} \protect\cite{hawthorn} \protect\cite{sutherland}.  In the d-wave
superconducting state the downturn is associated with a decrease in the nodal quasiparticle
contribution to $\kappa/T$ at a temperature well below that corresponding to the impurity scattering
rate.  This cannot be understood within standard dirty d-wave transport theory\protect\cite{durst}.

In this article we show how the observed downturns can be attributed to a decrease in the rate of
energy transferred from the phonons that provide the heat current to the electrons being studied
(from now on we will use the word ``electron'' to refer to either a Landau quasiparticle in the Fermi
liquid or a nodal Bogolubov quasiparticle in the d-wave superconductor).  As a result of the reduced
heat transfer rate, the phonons no longer come into thermal equilibrium with the electrons at very
low temperatures and the electronic heat current becomes dependent on the phonon-electron heat
transfer rate, and thus decreases rapidly with temperature.  With the
 origin of the downturn thus understood, the low-$T$ thermal conductivity data for cuprates in the superconducting state are consistent with the predictions of dirty-d-wave theory.  Similarly, the thermal conductivity of
normal state PCCO at very low temperatures exhibits a constant linear term in $\kappa/T$ versus $T$
down to the lowest temperatures, consistent with the behaviour measured above 300 mK, with no
downturn.  A detailed discussion of the experimental results in a variety of materials will appear in
Ref. \protect\onlinecite{collaboration_footnote}.

We develop a theoretical model of the temperature and field dependence of the downturn by calculating
the phonon-electron heat transfer rate and using it to determine the measured heat current and the
thermal conductivity.  In summary, our results (i) give the correct temperature dependence for the
observed thermal conductivity (ii) give the correct order of magnitude (which depends on the
electron-phonon matrix element) of the observed effect (iii) give the correct dependence of the
thermal conductivity on magnetic field in the vortex state.

In Section II of this article we describe the role that phonon-electron heat transfer plays in
low-temperature thermal conductivity measurements and explain the origin of the downturn.  In Section
III we show the calculation of the relevant electron-phonon heat transfer rate in the normal state
and in the vortex state of a d-wave superconductor as a function of temperature and magnetic field.
In Section IV we will compare our results with data.  In Section V we discuss some implications of
our results.

\section{The Effect of Phonon-Electron Heat Transfer on the Observed Thermal Conductivity at Low Temperature}

\begin{figure}
\includegraphics[width = 2.0 in, height = 2.0
in]{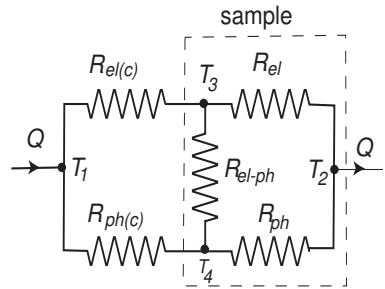} \caption{ \label{coupled_R}A simple picture of the experimental configuration for
thermal conductivity measurements.  The thermal resistance to phonon and electron heat currents that
occurs before the current enters the sample are represented by $R_{ph(c)}$ and $R_{el(c)}$,
respectively.  The thermal resistance to phonon and electron heat flow through the sample are
represented by $R_{ph}$ and $R_{el}$, respectively.The electron-phonon heat transfer rate is
associated with $R_{el-ph}$.  Temperatures at different positions have been labelled.}
\end{figure}

In order to study the effect of phonon-electron heat transfer rate on the
 low-$T$ thermal conductivity data, we use a simplified model (Fig. \ref{coupled_R}) of the experimental
  configuration.  A thermal current, $Q$, is carried into the sample by phonons and electrons.
The ``contact'' resistors, labelled $R_{el(c)}$ and $R_{ph(c)}$, correspond to the thermal resistance
encountered by the electronic and phononic heat currents before entering the sample.  Within the
cuprate material, the thermal resistance to the electron heat current, $R_{el}$, and to the phonon
heat current $R_{ph}$ are assumed to be described by the standard theory, i.e. by Fermi liquid
transport theory for the normal state and dirty-d-wave transport theory for the superconducting
state. The temperature and field dependent phonon-electron thermal resistance $R_{el-ph}$, describes
the heat transfer between phonons and electrons in the sample.

The model of Fig. \ref{coupled_R} is clearly a crude approximation of the actual experimental configuration, but
it turns out to be sufficient to describe the effect of phonon-electron heat transfer on the low-$T$
thermal conductivity measurements.  We will here briefly mention some of the differences between the
model shown and the actual experiment.  According to Fig. \ref{coupled_R}, the temperature difference
between phonons and electrons exists only at the hot end of the sample and the transfer of heat
between phonons and electrons occurs before either the phonon or electron heat currents encounter any
resistance from the sample.  In reality, a temperature difference between phonons and electrons can
be present everywhere in the sample and thus heat transfer can occur throughout the sample volume.
The temperature difference that drives the heat current through $R_{el-ph}$ actually depends on
position within the sample. However, since all our results below are calculated to linear order in
the temperature difference between electrons and phonons, we can determine the total phonon-electron
heat transfer rate below as if a constant temperature difference existed throughout the sample volume.  Also, in Fig. \ref{coupled_R} we have shown only one set of
current contacts, at the hot end of the sample, although a similar set of contacts must be present at
the cold end of sample in the experimental configuration. One could slightly improve the model of
Fig. \ref{coupled_R} by including a second, equivalent, set of resistors: $R_{el(c)}$, $R_{ph(c)}$
and $R_{el-ph}$ at the cold end of the sample (the right side of the figure).  This would account for
effects of the thermal resistance encountered by the phonon and electron heat currents after leaving
the sample and would allow for a temperature difference and resulting heat transfer between electrons
and phonons to occur at the cold end of the sample as well as the hot end.  We have determined the thermal conductivity for this modified model and have found that the effect of the modification to our main result below (Eq. \ref{conduct_modified}) is the same as
that of simply doubling both the value of $R_{el(c)}$ and the effective sample length over which phonon-electron heat transfer occurs.  Since we can only make rough comparisons of
$R_{el(c)}$ (as we discuss in Section IV) and the sample length (see footnote, Ref. \onlinecite{deltaT_footnote}) with measured quantities, this modification does not significantly change any of our results below.  We include these factors of two in comparisons with data below but will avoid needlessly cluttering the figure shown by including additional resistors.

As indicated in the introduction, the electron-phonon thermal resistance $R_{el-ph}$ plays a central
role in the determination of the low-$T$ behaviour of the apparent electronic contribution to the
thermal conductivity. In Section III, we calculate the electron-phonon thermal resistance for both
the normal metallic state and the d-wave superconducting state and find that the result in each case
is of the following form:
\begin{equation}
\label{el_ph_resistance} R_{el-ph}^{-1}=\Omega KT^n
\end{equation}
where $\Omega$ is the sample volume, $K$ is a constant, which depends on the electron-phonon matrix
element, that we determine and $n$ is between 4 and 5. For the remainder of this section, we will use
Eq. \ref{el_ph_resistance} along with Fig. \ref{coupled_R} in order to determine the low-$T$
behaviour of the apparent electronic thermal conductivity. This is done to provide a general
explanation for the observed low-$T$ downturns in both the normal state and the d-wave
superconducting state before getting into the details of the calculation of $R_{el-ph}$ for these
separate cases.

According to Eq. \ref{el_ph_resistance}, $R_{el-ph}$ varies more rapidly with temperature than any of
the other resistors in Fig. \ref{coupled_R}.  At sufficiently high temperatures, $R_{el-ph}$ is
negligible and the phonons and electrons at the hot end of the sample are in thermal equilibrium, so
$T_3=T_4$.  The thermal conductivity $\kappa$ can then be obtained from the thermal conductance $G$,
where $G=Q/(T_4-T_2)=R_{el}^{-1}+R_{ph}^{-1}$.  The result for either the Fermi liquid or the
dirty-d-wave superconductor using the scattering rates dominant at low-$T$ then has the usual form
\begin{equation}
\kappa/T=\alpha +\beta T^{2} \;\;\; (T>>T_D) \label{conduct}
\end{equation}
where the $\alpha T$ term is associated with the electron heat current and the $\beta T^{3}$ term is
associated with the phonon heat current.  We have introduced the characteristic downturn temperature
$T_D$, which corresponds to the temperature at which $R_{el-ph}=R_{el}$, that is the relevant
temperature scale for the downturn in the electronic contribution to $\kappa$.

At very low temperatures, $R_{el-ph}$ becomes large, and the phonons and electrons in the sample
become thermally decoupled.  There arises a temperature difference between the electrons and the
phonons at the hot end of the sample, $T_3\neq T_4$, that depends on the relative magnitude of
$R_{el(c)}$ and $R_{ph(c)}$.  There is thus some ambiguity in the definition of the thermal
conductivity.  A thermometer at the hot end of the sample could measure $T_3$, $T_4$, or some
function of the two, say $T^{\prime}=f(T_3,T_4)$ depending on its relative sensitivity to phonons and
to electrons.  The thermal conductivity at low-temperature would be determined from a conductance
given by $G=Q/(T^\prime-T_2)$.  One could consider many possible cases\protect\cite{phonon_bottle}
resulting from different forms of $f(T_3,T_4)$ over the relevant temperature range.  Here we focus on
one limiting case, $f(T_3,T_4)=T_4$ corresponding to a thermometer sensitive only to phonon
properties.  In this limiting case, an exact analysis of the configuration of Fig. \ref{coupled_R}
gives the conductance
\begin{equation}
\label{exact}
G\equiv\frac{Q}{T_4-T_2}=\frac{1}{R_{el}}\bigg{(}\frac{1+\frac{R_{ph(c)}}{R_{ph}}\Lambda}{1+\frac{R_{el(c)}}{R_{el}}\Lambda}\bigg{)}+\frac{1}{R_{ph}}
\end{equation}
where
\begin{equation}
\Lambda=\frac{R_{el-ph}}{R_{el-ph}+R_{el(c)}+R_{ph(c)}}. \nonumber
\end{equation}

In this article we do not try to understand the nature of the contact resistances $R_{el(c)}$ and
$R_{ph(c)}$ by looking into the details of the experiments, but instead determine what relative
values these quantities must have in order for us to be able to account for the observed
low-temperature thermal conductivity downturns.  To this end, note that when $R_{el(c)}$ becomes
infinite, one finds the thermal conductance
\begin{equation}
\label{zerophononr} \frac{Q}{T_4-T_2}=\frac{1}{R_{el}+R_{el-ph}}+\frac{1}{R_{ph}}
\end{equation}
where again we have assumed that the hot end of the sample measures the phonon temperature $T_4$.  At
$T>>T_D$ where $R_{el-ph}<<R_{el}$, Eq. \ref{zerophononr} gives Eq. \ref{conduct} above.  However, at
temperatures $T<<T_D$, where $R_{el-ph}$ describes a high resistance to heat flow between phonons and
electrons, the effective electronic contribution to the heat conductivity (the first term on the
right side of Eq. \ref{zerophononr}) is much reduced.  This is our picture of the low-temperature
downturns.  Alternatively, if the value of $R_{ph(c)}$ is taken to be infinite, the determined
thermal conductivity shows no low-$T$ downturn (this situation is discussed briefly in Ref.
\protect\onlinecite{phonon_bottle}).  To retain the simplest model of the experimental configuration
that gives the correct low-$T$ behaviour, we can take $R_{el(c)}>>R_{ph(c)}$ over the entire
temperature-range considered\protect\cite{alternative_limit}.

Following this logic, we neglect terms of order $R_{ph(c)}/R_{el(c)}$ compared to unity in Eq.
\ref{exact}.  We then obtain, after some rearrangement and the use of Eqs. \ref{el_ph_resistance} and
\ref{conduct} the thermal conductivity given by
\begin{equation} \kappa/T=\alpha \frac{1}{1+\frac{r}{1+r(T/T_D)^{n-1}}}+\beta T^2
\label{conduct_modified}
\end{equation}
where $T_D\equiv (\alpha /K l_s^2)^{1/(n-1)}$, $l_s$ is the length of the sample along the direction
of the current and $r\equiv R_{el(c)}/R_{el}$ will be assumed to be independent of temperature. This
is a generalization of Eq. \ref{zerophononr} valid for finite $R_{el(c)}$. At temperatures much
larger than $T_D$ the phonons and electrons in the sample are in local thermal equilibrium and Eq.
\ref{conduct_modified} reduces to Eq. \ref{conduct}. At temperatures well below $T_D$ the electronic
contribution is limited by phonon-electron heat transfer, so that $\kappa\propto R_{el-ph}^{-1}$ due
to the presence of the heat current bottleneck. Thus the electronic contribution decreases to its
minimum value, $\kappa_0=\alpha T/(1+r)$, as $\kappa-\kappa_0 \propto (T/T_D)^n$ at low temperature.

Eq. \ref{conduct_modified}, along with its field-dependent generalization (Eq. \ref{conduct_field})
that applies in the vortex state, is the main result of this article.  If we use the values of $n$
and $T_D$ that we calculate in Section III, then Eq. \ref{conduct_modified} gives a good description
of the low-$T$ data for downturns observed in the normal and superconducting states of cuprates.

\section{Calculation of the phonon-electron heat transfer
rate in the normal and superconducting state}

In considering the electron-phonon thermal decoupling, we assume that the electrons in the sample
remain in local thermal equilibrium at a certain temperature $T_{el}$.  We also assume that the
phonons maintain local thermal equilibrium at a temperature $T_{ph}$.  We consider the possibility
that the local electron temperature $T_{el}$ differs from the local phonon temperature, $T_{ph}$ and
determine the resulting rate of heat transfer from the hot phonons to the cold electrons.  This
transfer rate is given by \protect\cite{kaga01}

\begin{equation} \frac{dU}{dt} =  \sum_{\bf q,j} \hbar \omega
_{\bf q,j} \alpha_{\bf{q}j}c_{\bf{q}j}
   \bigg[ n(\frac{\hbar \omega _{{\bf q},j}}{k_{B}T_{ph}})-n(\frac{\hbar
\omega_{{\bf
   q},j}}{k_{B}T_{el}}) \bigg]
\label{dudt1} \end{equation} where the sum is taken over phonons of wavevector ${\bf q}$, energy
$\hbar\omega_{{\bf q},j}$, velocity $c_{\bf{q},j}$, and polarization $j$ and $n(x)$ is a Bose factor.
All the electronic properties are contained within the sound attenuation $\alpha_{\bf{q}j}$ which is
calculated separately for the normal and superconducting states.

The difference between the local temperatures of phonons and electrons is at most equal to the
observed temperature difference between the two ends of the sample $(T^{\prime}-T_2)$, which
corresponds to the entire range of temperature in the system. Since this temperature range is much
smaller than the average temperature for these experiments, we can expand the above expression to
first order in $\Delta T=T_{ph}-T_{el}$, so that the heat transfer rate can be expressed as a
electron-phonon thermal resistance

\begin{equation}
\frac{dU}{dt}=\Delta T/R_{el-ph}(T) \label{Rdef}
\end{equation}
where $T$ is the average temperature of the sample and $\Delta T$ can be thought of as the average
temperature difference between the phonons and electrons over the sample
volume\protect\cite{deltaT_footnote}.

The fact that the thermal conductivity in PCCO is independent of field \protect\cite{normal_state}
over the range from 8-13T suggests that the material is in the dirty limit, i.e. $\omega_C\tau<<1$
where $\tau^{-1}$ is the scattering rate due to impurities and $\omega_C$ is the cyclotron frequency.
This suggestion is supported by the electrical conductivity data, which show a residual resistivity
corresponding to $\hbar\tau^{-1}/k_B\approx$130K, which gives $\omega_C\tau=$0.1 for the highest
field reported.  The field has no effect on electronic transport in this limit, and we ignore it in
our calculation of $R_{el-ph}$ for this material.  The condition for the low-temperature limit of the
heat transfer rate: $\hbar\tau^{-1}>>(v_f/c_s)k_BT$ (which corresponds to the low-frequency limit of
sound attenuation for thermally excited phonons) is
satisfied\protect\cite{measuredRelph}\protect\cite{gersh}\protect\cite{roukes}\protect\cite{anderson}.
With this, the well known result for the sound attenuation in metals in the low-frequency limit is
appropriate for PCCO. The normal state attenuation is given by
\begin{equation}
\alpha^N_{\bf{q}j}c_{\bf{q}j}=2\tau_Nn_0(\hbar\omega_{{\bf
q}j})^2\frac{g^2}{Mc_{\bf{q}j}^2}<|f_j({\bf k},{\bf q})|^2>_{FS} \label{alphaN}
\end{equation} where
$\tau_N^{-1}$ is the impurity scattering rate obtained from the residual electrical resistance, $n_0$
is the density of states and $M$ is the mass of a unit cell. The vector subscripts on the symbol,
$\alpha^N_{\bf{q}j}$ for the sound attenuation should remove possible confusion with the $\alpha$
used in Eq. \ref{conduct} to represent the electronic term in the thermal conductivity, so we will
continue to use this standard notation.  The electron-phonon interaction is described as in Ref.
\protect\onlinecite{walk01}, where $g^2$ is the coupling constant that sets the energy scale for the
interaction and $f_j(\bf{k},\bf{q})$ is the dimensionless factor which describes directional and
phonon-mode dependence.  The explicit form of $f_j(\bf{k},\bf{q})$ for each high-symmetry phonon mode
is given in Ref. \onlinecite{walk01}.

We determine the thermal resistance using Eqs. \ref{dudt1}, \ref{Rdef} and \ref{alphaN}.
Substituting Eq. \ref{alphaN} into Eq. \ref{dudt1} and using
\begin{equation}
\label{bose_ex}
 n(\frac{\hbar \omega _{{\bf q},j}}{k_{B}T_{ph}})-n(\frac{\hbar
\omega_{{\bf
   q},j}}{k_{B}T_{el}})=-\frac{\omega_{{\bf q},j}}{k_BT}\bigg{[}\frac{\partial n(\frac{\hbar \omega _{{\bf q},j}}{k_{B}T})}{\partial \omega_{{\bf q},j}}\bigg{]}k_B\Delta T,
\end{equation}
we obtain
\begin{equation} \label{dudt_final}\begin{split}\frac{dU}{dt} = k_B &\Delta T (k_BT)^5\bigg{[}\frac{2\tau_Nn_0 g^2}{\rho(2\pi\hbar)^3}\bigg{]} \\&\sum_{j}\int d\Omega_{\bf q}\frac{<|f_j({\bf
k},{\bf q})|^2>_{FS}}{c_{{\bf q}j}^5} \int dx x^6 \big{(}-\frac{d n}{dx}\big{)}
\end{split}
\end{equation}
where $\rho$ is the mass density and the $\Omega_{\bf q}$ integral is over all phonon directions
(both $f_j({\bf k},{\bf q})$ and $c_{{\bf q}j}$ depend on the mode and direction of the phonon but
not on its energy). 

The electron-phonon heat transfer rate in the normal state is thus seen to be
proportional to $T^5$ at temperatures that satisfy $\hbar\tau^{-1}>>(v_f/c_s)k_BT$.  This dependence can be understood by power counting as follows: two factors of temperature come from the $\omega_{{\bf q},j}^2$ dependence of the sound attenuation, three from the phonon density and an additional one from the energy of the phonon that appears in Eq. \ref{dudt1} with one power of temperature removed by the factor $\Delta T/T$ in Eq. \ref{bose_ex} that describes the balance of energy exchange.  We note that for the
temperature $T$ in Eqs. \ref{bose_ex} and \ref{dudt_final} we can use either $T_{ph}$ or $T_{el}$ since these expressions are already first order in $\Delta T$.

By comparing Eq. \ref{Rdef} with Eq. \ref{dudt_final} we obtain $R_{el-ph}^{-1}$, which is
proportional to $T^5$, and use this along with the definition of $T_D$ given in the Introduction to
determine the downturn temperature.  The result is
\begin{equation}
\label{TD_value}
 T_{D}^{-4}=\frac{l_s^2}{\alpha} \bigg{(}\frac{8 \pi \Gamma(7) \zeta(6) \tau_N n_0
k_B^6 g^2 \langle\langle|f_j({\bf k},{\bf q})|^2 c^{-5}_{\bf{q}j}\rangle\rangle}{(2 \pi \hbar)^3 M}
\bigg{)}
\end{equation}
where the $\langle\langle x\rangle\rangle$ indicates that the electron-phonon interaction has been
averaged over the Fermi surface and phonon direction and summed over phonon modes.  For the numerical
estimates of $T_D$ described in Section IV we approximate the phonon averages by using the known
expressions of $f_j({\bf k},{\bf q})$ for high-symmetry directions and interpolating them to all
other directions.  The sound velocity of both transverse and longitudinal modes is taken to be
isotropic.  Transverse modes have a sound velocity that is significantly smaller than that for
longitudinal modes, which results in a heavier weight of the contribution by the former to
$T_D^{-4}$.

In the d-wave superconducting state, the low-temperature limit of $R_{el-ph}$, with its
characteristic $T^5$ dependence is valid at temperatures such that $\gamma>>(v_f/c_s)k_BT$, where
$\hbar^{-1}\gamma$ is the zero-frequency scattering rate of nodal quasiparticles by impurities.  In
this limit, the sound attenuation and hence $R_{el-ph}$ are independent of $\gamma$, which is an
example of the universal behaviour of transport properties in d-wave
superconductors\protect\cite{lee}\protect\cite{tail01}. The sound attenuation in this limit is
related to the normal state expression of Eq. \ref{alphaN} by

\begin{equation}
\label{attenuation_universal} \alpha^S_{\bf{q}j}c_{\bf{q}j}=(\tau_N^{-1}/\pi
v_2k_n)\alpha^N_{\bf{q}j}c_{\bf{q}j}\frac{|f_j({\bf k}_{n},{\bf q})|^2}{<|f_j({\bf k},{\bf
q})|^2>_{FS}}
\end{equation}
where ${\bf k}_n$ is the length of the wavevector to a node, and $\hbar v_2$ is the slope of the gap
at the node.  We can use Eq. \ref{attenuation_universal} in Eq. \ref{dudt1} to obtain the $T^5$
dependence of the electron-phonon heat transfer rate (note that the d-wave density of states is constant at temperatures $k_BT\ll\gamma$ so that the power counting proceeds just as for the normal state).  The value of $T_D^{-4}$ for the
superconducting state is obtained by multiplying the right side of Eq. \ref{TD_value} by the factor
$\alpha^S_{\bf{q}j}c_{\bf{q}j}/\alpha^N_{\bf{q}j}c_{\bf{q}j}$ given by Eq.
\ref{attenuation_universal}.  In the resulting expression for $T_D^{-4}$, the matrix element
$|f_j({\bf k}_{n},{\bf q})|^2$ is to be averaged over phonon direction and summed over phonon modes.

Typical experimental values of $\gamma$ (which is related to $\tau_N$ by
$\gamma\approx0.61\sqrt{\Delta_0 \hbar \tau_N^{-1}}$ if we assume unitary scattering)
\protect\cite{durst} for most clean cuprate samples suggest that the strict low-$T$ limit of
$R_{el-ph}$ occurs below the lowest temperatures used for thermal transport measurements.  We thus
must determine $R^{-1}_{el-ph}$, which becomes weakly dependent on $\gamma$, numerically over the
experimental temperature range.  The result of such a computation, using realistic values of $\gamma$
corresponding to between 1-15K as estimated from the position of the peak in $\kappa$ as a function
of temperature below $T_C$ (see Ref. \protect\onlinecite{sutherland} and references therein), is
well-described by a power law with an exponent $n$ between 4 and 5.  For example, by choosing
$\gamma$ to correspond to a temperature of 3K, we find an exponent of $n=4.4$, which can be used in
Eq. \ref{conduct_modified}.  When we compare our results to the in-field data on LSCO in Section IV,
we will use the low-$T$ (universal limit) result in order to simplify the comparison with the normal
state result and the vortex state calculation described below.  This result is only strictly valid
within the rather narrow parameter range of $\hbar(v_f/c_s)k_BT<<\gamma<<\Delta_0$, but should give a
reasonable description for the case of LSCO over the temperatures of interest.  If we were to allow
$\gamma$ to vary over its experimental range, then we could adjust the value of $n$ somewhat and
perhaps improve the fit to the data.  However, as seen in Section IV, the agreement between universal
limit result and the available data is already very good.  Thus in what follows we will use the
universal limit result and will take $k_B^{-1}\gamma=11$K for LSCO (the zero-field result is
independent of $\gamma$ but $\gamma$ does appear in the field-dependent calculation below) in
agreement with some estimates for this material from Ref. \protect\onlinecite{sutherland}.

To study the magnetic field dependence of downturns observed in the vortex state of cuprates with a
field applied perpendicular to the CuO$_2$ planes we follow the approach of Volovik,
 and replace the frequency appearing in the superconducting Green's
function with a Doppler-shifted value\cite{volovik}.  The doppler shift is given by $\hbar{\bf v}_s\cdot{\bf k}_n$
where $v_s$ is the supercurrent velocity which depends on the position within the vortex-lattice unit
cell, and ${\bf k}_n$ is a wavevector directed to the d-wave node.  The nodal wavevector can be used
instead of the true electron momentum at the low temperatures of interest.  The phonon-electron
energy transfer rate can then be obtained as a position average over a simplified vortex-lattice unit
cell\protect\cite{kubert}.  Analytic results can be obtained in the clean, low-$T$ limit defined by
$\hbar{\bf v}_s\cdot{\bf k}_n>>\gamma, (v_f/c_s)k_BT$, which can be expected to be valid in fields of
several Tesla in many cuprates (here we refer to a typical value of ${\bf v}_s\cdot{\bf k}_n$ within
the vortex-lattice unit cell).  The formal similarity between the expression for the phonon
attenuation rate and the thermal conductivity results in an identical field dependence for the
electron-phonon heat transfer rate and the heat conductivity. The field-dependent attenuation in the
superconducting state is given by
\begin{equation}
\alpha^S_{\bf{q}j}(H)=({\bf v}_s\cdot{\bf
k}_n)^2\big{(}\frac{\pi}{\tau_N^{-1}v_2k_n}\big{)}\alpha^S_{\bf{q}j}(0)\label{alphaS}
\end{equation}
where $v_2$ is the magnitude of the slope of the superconducting gap at the node and
$\alpha^S_{\bf{q}j}(0)$ is given by Eq. \ref{attenuation_universal}.  The expression for the
superconducting impurity scattering rate in the clean, unitary limit
$\gamma=\tau_N^{-1}\Delta_0/2\bf{v}_s\cdot\bf{k}_n$ has been used.  The electronic thermal
conductivity is given in the same limit by
\begin{equation}
\frac{\kappa_{el} (H)}{T}=({\bf v}_s\cdot{\bf k}_n)^2\big{(}\frac{\pi}{\tau_N^{-1}
v_2k_n}\big{)}\frac{\kappa_{el}(0)}{T} \label{kclean}
\end{equation}
Eq.'s \ref{alphaS} and \ref{kclean} are to be averaged over a unit cell of the vortex lattice.  We
follow Ref. \protect\onlinecite{kubert} in performing the vortex unit cell average appropriate for
fields applied perpendicular to the planes and assume unitary scattering.  Both the thermal
conductivity and the sound attenuation (and consequently $R^{-1}_{el-ph}$) are thus found to be
proportional to $\sqrt{H/H_{C2}}$.  If we add the zero field universal result we have a decent
description of the field dependences of $\alpha(H)$ and $R^{-1}_{el-ph}$ for applied fields up to
some fraction of $H_{C2}$.  We then substitute into Eq. \ref{conduct_modified} to obtain
\begin{equation}
\kappa_{el}/T=\frac{\eta(H)\alpha}{1+\frac{\eta(H)r}{1+\eta(H)r(T/T_D)^{n-1}}} \label{conduct_field}
\end{equation}
with
\begin{equation}
\eta(H)=1+\sqrt{\frac{H}{H_{C2}}}\frac{\alpha^{\prime}}{\alpha}
\end{equation}
where $\alpha^{\prime}=a(3\pi/8)\sqrt{\Delta_0\hbar^{-1}\tau_N}\alpha$, and $a$ is a factor of order
unity that arises because we roughly approximated the vortex lattice geometry.  The field dependence
of the expression is contained entirely in $\eta(H)$.  The factors $r$ and $T_D$ are to be evaluated
at $H=0$.

\section{Comparison with the data}

\begin{figure} \includegraphics[width = 2.7 in, height = 2.16
in]{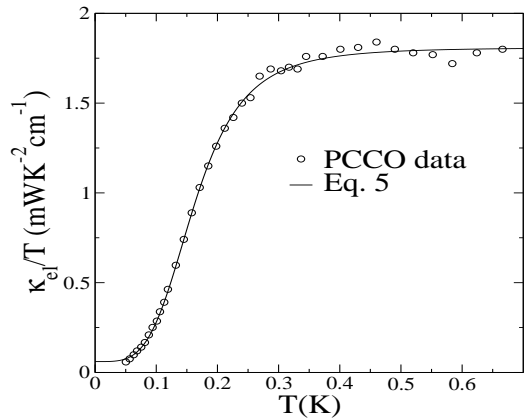} \caption{ \label{pcco_data} The electronic thermal conductivity
$\kappa_{el}/T$ measured \protect\cite{rhill} for normal state Pr$_{2-x}$Ce$_x$CuO$_{7-\delta}$ is
plotted versus $T$ along with the result of Eq. \ref{conduct_modified}.}
\end{figure}

\begin{figure} \includegraphics[width = 2.7 in, height = 2.16
in]{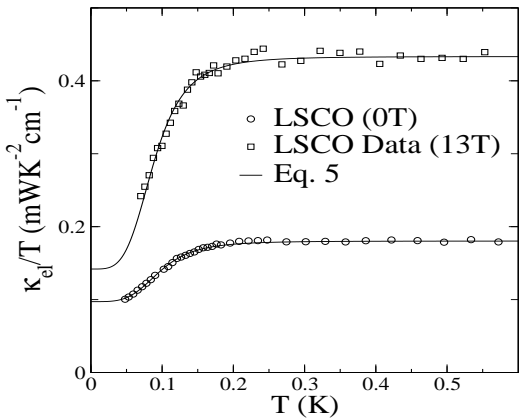} \caption{ \label{lsco_data} The electronic thermal conductivity
$\kappa_{el}/T$ versus $T$ measured\protect\cite{hawthorn} for La$_{2-x}$Sr$_x$CuO$_4$ in the
superconducting state in zero field and in a field of 13T applied perpendicular to the CuO$_2$ planes
along with the results of Eq. \ref{conduct_modified}.}
\end{figure}

In Fig. \ref{pcco_data} and Fig. \ref{lsco_data} we show data for the field-induced normal state of
optimally doped PCCO\protect\cite{rhill} and for superconducting, optimally doped
LSCO\protect\cite{hawthorn}, respectively.  The plots are of the extracted electronic contribution to
$\kappa/T$ versus $T$.  Thus the phonon contribution has been subtracted from the data to leave only
that of the electrons.

According to our discussion in Section II, the phonon contribution could be identified by fitting the
data at temperatures well above $T_{D}$ to Eq. \ref{conduct_modified}. In practice, the second term
in Eq. \ref{conduct_modified} often does not adequately describe the phonon contribution (for
example, slightly weaker temperature power-laws are often needed as discussed in Ref.
\protect\onlinecite{sutherland}). For our purposes, as long as the phonon contribution is determined
from data at temperatures well above $T_D$, then once it has been subtracted we can use our results
above and consider only the first term in Eq. \ref{conduct_modified}.  In Ref.
\protect\onlinecite{rhill}, the authors describe in detail how the phonon subtraction was performed
for the PCCO data.  For the LSCO data, we performed the phonon subtraction by first fitting the data
of Ref. \protect\onlinecite{hawthorn} at temperatures above the downturn (and below the maximum
temperature shown on Fig. \ref{lsco_data}) to the form $\kappa/T=\alpha+\beta T^m$ where $m$ was
between 1 and 2 and then subtracting the $\beta T^m$ term.  For both the LSCO data and the PCCO data,
the electron contribution to $\kappa/T$ is seen to be independent of temperature for $T>>T_D$.

\begin{figure} \includegraphics[width = 2.7 in, height = 2.16
in]{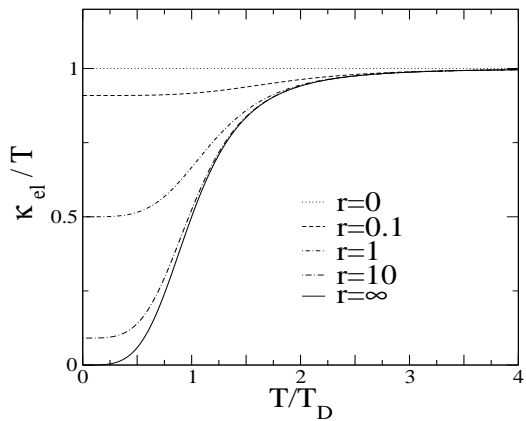} \caption{ \label{versusr} $\kappa_{el}/T$, normalized to its high-$T$
value, determined by Eq. \ref{conduct_modified} is plotted versus $T$ for various values of the
``contact'' resistance $r$.}
\end{figure}

We compare the result of Eq. \ref{conduct_modified} in the normal state to the PCCO data in Fig.
\ref{pcco_data}.  For the result shown we used the parameters: $\alpha=1.75$mWK$^{-2}$cm$^{-1}$ that
we take from the data at temperatures above the downturn, $n=5$ that we calculated in Section III,
 $T_{D}=160$mK, which was chosen to fit the data and $r=29$ that was chosen to fit the data.  Evidently Eq. \ref{conduct_modified} captures the
low-$T$ behaviour of the data.  The large value of $r$ indicates that heat is carried into the sample
almost exclusively by phonons.  Indeed, taking $r=\infty$ gives a reasonably good description of the
PCCO data.  This can be seen from Fig. \ref{versusr}, in which we show the result of Eq.
\ref{conduct_modified} for various values of $r$.

In order to determine whether the value of $T_D=160$ mK is realistic for this material, we can
compare this value with an estimate based on Eq. \ref{TD_value}.  All the parameters in Eq.
\ref{TD_value} are known for cuprates except the electron-phonon coupling constant, $g^2$.  The value
of $g^2$ has been obtained from sound attenuation measurements\protect\cite{lupien} on Sr$_2$RuO$_4$,
another material to which the given form of the electron-phonon interaction is applicable, and found
as $g^2=(12$eV$)^2$. Although the electron-phonon coupling constant may vary significantly for
different materials, we can expect order-of-magnitude agreement between the value of $g^2$ for
Sr$_2$RuO$_4$ and that for PCCO.  Using Eq. \ref{TD_value} and the value $T_D=160$ mK, we calculate
the electron phonon coupling constant for PCCO and find $g^2=(6$eV$)^2$.  Thus, the coupling
constant for PCCO determined using the fit above roughly agrees with the value measured for
Sr$_2$RuO$_4$.

In Fig. \ref{lsco_data} the result of Eq. \ref{conduct_modified} for the case of the d-wave
superconductor in the low-temperature limit as a function of applied magnetic field is compared to
the LSCO data.  Here, the zero-field result was obtained using the values:
$\alpha=0.18$mWK$^{-2}$cm$^{-1}$, $n=5$, $T_{D}=80$mK, $r=0.86$ that were chosen exactly as for the
PCCO case described above.  In order to obtain the 13T result, we use Eq. \ref{conduct_field} with
the values of $\alpha$, $n$, $T_D$, and $r$ fixed by the zero-field fit (the two curves in Fig. \ref{lsco_data} were measured for the same sample).  The remaining unknown
factor of order unity, $a$, that appears in Eq. \ref{conduct_field} is obtained from the 13T data at
temperatures well above the downturn (we thus find $a=0.5$).  With this done, the low-temperature
downturn for the 13T curve is determined without any adjustable parameters.  The clear agreement with
the data thus provides strong support for our model.

To check whether the value of $T_D=80$ mK for superconducting LSCO is reasonable we compare it with
the value determined in Section III.  We note that the electronic thermal conductivity in the
universal limit is related to the normal state Boltzmann expression by an expression similar to Eq.
\ref{attenuation_universal}, namely $\kappa^S_0/T=\kappa^N_0/T (\tau_N^{-1}/\pi v_2k_n)$.  As a
result, the value of $T_D$ determined using the low-$T$ results for both $R_{el-ph}$ and
$\alpha=\kappa_0/T$ is roughly the same in the superconducting state of cuprates as it was for the
normal state.  From Eq. \ref{TD_value} (also, recalling the discussion immediately following Eq.
\ref{attenuation_universal}) we find that the value of $T_D=80$mK corresponds to $g^2=(7$eV$)^2$.
The agreement of the electron-phonon coupling constants for PCCO and LSCO is expected since $g^2$ is
a normal state property of the CuO$_2$ planes.

\begin{figure} \includegraphics[width = 2.7 in, height = 2.16
in]{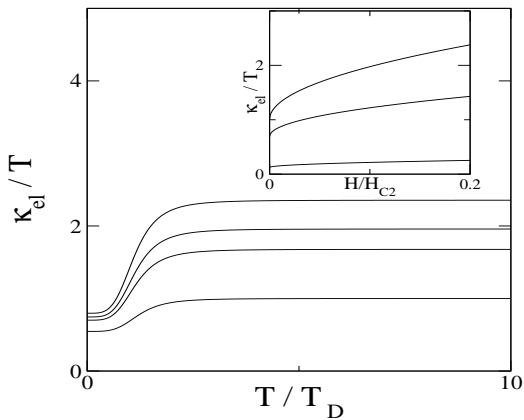} \caption{ \label{versush} Main Panel: $\kappa_{el}/T$, normalized to its high$-T$ zero-field value, versus $T$ in
varying magnetic fields as given by Eq. \ref{conduct_field}.  From the bottom, the curves are for
fields of 0, 0.05, 0.1, and 0.2 $H_{C2}$ applied perpendicular to the CuO$_2$ planes. Inset: The same quantity
plotted versus $H$ at various temperatures.  From the bottom, the curves are for temperatures of 0.1,
1, and 10 $T_D$.}
\end{figure}

The parameter $r\equiv R_{el(c)}/R_{el}$ determines the extent of the low-$T$ downturn, i.e. the
value of $\kappa/T$ at $T=0$. The value of $r$ corresponds to the thermal resistance to heat being
transmitted to the electrons in the sample by any means other than via phonon-electron heating within
the sample.  We have not attempted to describe this complicated process and have treated $r$ as an
unknown parameter. However, if we assume that the main contribution to $r$ comes from the resistance
of the current contacts, which is usually measured during thermal conductivity experiments, then we
can compare the values of $r$ obtained above to the measured values.  We convert the thermal
conductivity of the sample measured well above the downturn to an electrical resistance by using the
Wiedemann Franz law, according to which
\begin{equation}
R_{el}=\frac{l_s}{A}L_0\big{(}\frac{\kappa_{el}}{T}\big{)}^{-1}
\end{equation}
where $A$ is the cross-sectional area of the sample and $L_0$ is the Lorenz number.  For the PCCO
measurements, the contacts were reported \cite{rhill} to have a measured resistance of roughly 1
$\Omega$. So, using the reported sample dimensions, $l_s=1 $ mm, $A=38$ $\mu$m $\times 720 \mu$m and
using $R_{el(c)}=2 \Omega$ (the factor of two mentioned in the discussion following Fig. \ref{coupled_R} has been included) we get $r=240$.  This is substantially larger than the best-fit value of 29, but from Fig. \ref{versusr} is seen to still be reasonably consistent with the data.  For the LSCO data, the reported contact resistance was roughly 10 m$\Omega$. The
dimensions of the sample\cite{hawt_private} for which the data above was obtained were $l_s=0.97$ mm,
$A=1.26 $mm$\times 0.212$ mm. From this we obtain $r=1.8$, in rough agreement with the value $r=0.9$ used above.  We do not expect precise agreement between the $r$ obtained above and the measured contact resistance, but it is clear that the values of $r$ used in our fits are not out of line with the measured values.

The data shown above was chosen for being representative of that in which downturns are seen.  Other
examples of downturns, some of which were discussed in Section I, typically resemble those shown
above and occur at roughly the same temperature.

\section{Discussion}

We have shown that the low-T downturn in the electronic contribution to thermal conductivity, for
PCCO in the normal state and LSCO in the superconducting state, results from the loss of thermal
equilibrium between phonons and electrons caused by poor heat transfer between these systems at
low-temperature.  The electron-phonon heat transfer rate has been shown
 by calculation to have both a characteristic temperature dependence in the
 normal state, and a characteristic temperature dependence and magnetic field
 dependence in the superconducting state.
The fact that these dependences are in excellent agreement with experimental observations in PCCO and
LSCO and provides a striking confirmation that poor electron-phonon heat transfer at low temperatures
is indeed the right mechanism to explain the downturns in these
materials. Clearly this is a mechanism that should
also be considered in connection with
 thermal conductivity downturns in other materials\protect\cite{measure_248}\protect\cite{hussey}.  If one seeks to measure
the electronic contribution to $\kappa/T$ at zero temperature, the role of electron-phonon thermal
decoupling has to be carefully considered. Alternatively, one could study the downturns to extract
information about the electron-phonon interactions, without regard to the underlying thermal
conductivity.

Thermal conductivity data at temperatures below $T_D$ that shows a significant downturn could be used
to study electron-phonon interaction effects in unconventional superconductors and other systems of
interest. In d-wave superconductors, the electron-phonon heat transfer rate calculated within the
dirty d-wave approach has many of the same features as the thermal conductivity.  We showed that it
has the same magnetic field dependence at low-temperature, and also shows universal behaviour (albeit
at a temperature which is smaller by a factor of the ratio of the speed of sound to the fermi
velocity, $c_s/v_f$ than that at which universality occurs in the thermal
conductivity.\protect\cite{lee} \protect\cite{tail01} \protect\cite{smith}) At higher temperatures,
the field and temperature dependences of the heat transfer rate have different forms from those of
thermal conductivity, and provide complementary information about the electrons.  The basic form of
the electron-phonon matrix elements, which affects the heat transfer rate, also has unusual
properties in layered square-lattice tight-binding materials\protect\cite{walk01}
\protect\cite{lupien}.

\section{Acknowledgments} We acknowledge the support of the
Canadian Institute for Advanced Research and of the Natural Sciences and Engineering Research Council
of Canada.  We are grateful to D. G. Hawthorn, M. L. Sutherland and R. W. Hill for useful discussions
and for providing the data of Fig. \ref{pcco_data}.  This article has been published in Physical Review B, Ref. \protect\onlinecite{thisarticle}, and copyright is owned by the APS.

$\dagger$ Present address: National Synchrotron Research Center, Nakhon Ratchasima, Thailand.
$\dagger\dagger$ Present address: Department of Physics, University of California, San Diego, La Jolla, CA, USA. 
$\ddagger$ Members of the Quantum Materials Program of the Canadian Institute for Advanced Research.


\begin{thebibliography}{99}

\bibitem{rhill} R. W. Hill, Cyril Proust, Louis Taillefer, P.
Fournier and R. L. Greene, Nature {\bf 414}, 711 (2001).

\bibitem{zheng} Guo-qing Zheng, T. Sato, Y. Kitaoka, M. Fujita,
and K. Yamada, Phys. Rev. Lett. {\bf 90}, 197005 (2003).


\bibitem{nakamae} S. Nakamae, K. Behnia, N. Mangkorntong, M.
Nohara, H. Takagi, S. J. C. Yates, and N. E. Hussey, Phys. Rev. B {\bf 68}, 100502 (2003).

\bibitem{takeya} J. Takeya, Yoichi Ando, Seiki Komiya, and X. F.
Sun, Phys. Rev. Lett. {88}, 77001 (2002).

\bibitem{hawthorn} D. G. Hawthorn, R. W. Hill, C. Proust, F.
Ronning, Mike Sutherland, Etienne Boaknin, C. Lupien, M. A. Tanatar, Johnpierre
 Paglione, S. Wakimoto, H. Zhang, Louis Taillefer, T. Kimura, M. Nohara, H.
 Takagi and N. E. Hussey, Phys. Rev. Lett {\bf 90} 197004 (2003).

\bibitem{sutherland} M. Sutherland, D. G. Hawthorn, R. W.
Hill, F. Ronning, S. Wakimoto, H. Zhang, C. Proust, Etienne Boaknin, C. Lupien and Louis Taillefer,
Phys. Rev. B {\bf 67}, 174520 (2003).




\bibitem{durst} A. C. Durst and P. A. Lee, Phys. Rev. B {\bf 62},
1270 (2000).

\bibitem{collaboration_footnote} Louis Taillefer {\it et al} to be published.

\bibitem{deltaT_footnote} We have evidently treated the geometry very crudely, and in a more accurate
treatment would be concerned with the relative positions of the current contacts and the thermometers
on the sample and the effective sample volume over which phonon-electron heat transfer is relevant.  For example, in Eq. \ref{conduct_modified} the factor $l_s^2$ contains one factor of length coming from $R_{el}$ that should be equal to the distance between the thermometer contacts and one factor of length coming from $R_{el-ph}$ that should be equal to the effective length over which phonon-electron heat transfer is significant.  We ignore these details and use a single length parameter $l_s$. Since all the relevant distances (i.e. between contacts, from contact to sample ends etc.) are typically reasonable
fractions of the sample length, we expect that our result will be correct to within a factor of unity.  The value of the electron-phonon coupling constant $g^2$ is thus determined to within such a factor.

\bibitem{phonon_bottle} For example, in the opposite limiting case, when $f(T_3,T_4)=T_3$, the contribution of the phonons to the thermal current is reduced by an energy bottleneck at low-$T$.  There will be a low-$T$ upturn of $\kappa/T$ occurring at $T_U=(\beta/Kl_s^2)^{\frac{1}{n-3}}$.

\bibitem{alternative_limit} The same result, Eq. \ref{conduct_modified}, can be obtained by taking the combined limits $R_{ph(c)}<<\max(R_{el(c)},R_{el-ph})$ and $R_{ph(c)}<<R_{ph}$.

\bibitem{kaga01} M. I. Kaganov, I. M. Lifshitz and L. V.
Tanatarov Sov. Phys. JETP, {\bf 4}, 173 (1957).

\bibitem{normal_state} Hill {\it et al} determined that the sample was in its normal state for fields above 8T.

\bibitem{measuredRelph} $R_{el-ph}$ has been measured in disordered metallic films in the low-temperature limit and found to have the predicted $T^5$ temperature-dependence (Ref. \protect\onlinecite{gersh}).  In measurements of clean metallic samples in the opposite (high-temperature) limit, the predicted $T^4$ dependence of $R_{el-ph}$ has been observed (Refs. \protect\onlinecite{roukes} and \protect\onlinecite{anderson}).

\bibitem{gersh} M. E. Gershenson, D. Gong, T. Sato, B. B. Karasik, A. V. Sergeev, App. Phys. Lett. {\bf 79}, 2040 (2001).

\bibitem{roukes} M. L. Roukes, M. R. Freeman, R. S. Germain, R. C. Richardson, and M. B. Ketchen Phys. Rev. Lett. {\bf 55}, 422 (1985).

\bibitem{anderson} A. C. Anderson and R. E. Peterson, Phys. Lett. {\bf 38A}, 519 (1972).

\bibitem{walk01} M.B. Walker, M. F. Smith, and K. V. Samokhin,
Phys. Rev. B {\bf 65}, 014517 (2002).

\bibitem{lee} P. A. Lee, Phys. Rev. Lett. {\bf 71}, 1887 (1993).

\bibitem{tail01} L. Taillefer, B. Lussier, R. Gagnon, K. Behnia,
and H. Aubin, Phys. Rev. Lett. {\bf 79},483(1997).

\bibitem{volovik} G. E. Volovik, JETP Lett. {\bf 58}, 469 (1993).

\bibitem{kubert} C. K\"{u}bert and P. J. Hirschfeld, Phys. Rev.
Lett. {\bf 80} 4963 (1998).

\bibitem{lupien} C. Lupien, W. A. MacFarlane, Cyril Proust, Louis
Taillefer, Z. Q. Mao, and Y. Maeno, Phys. Rev. Lett. {\bf 86}, 5986 (2001).

\bibitem{hawt_private} D. G. Hawthorn, Private Communication.

\bibitem{measure_248} In measurements made on superconducting
YBa$_2$Cu$_4$O$_8$ (Ref.\protect\onlinecite{hussey}), the thermal conductivity data showed no
downturn with decreasing temperature, but the extracted $T$-linear term in the conductivity was
negligible.  This is likely another example of electron-phonon thermal-decoupling. The low-$T$
electronic heat current may be reduced, but the $T$-dependent downturn obscured by the subtraction of
the phonon contribution.

\bibitem{hussey} N. E. Hussey, S. Nakamae, K. Behnia, H. Takagi,
C. Urano, S. Adachi, S. Tajima, Phys. Rev. Lett {\bf 85}, 4140 (2000).

\bibitem{smith} M. F. Smith and M. B. Walker, Phys. Rev. B {\bf 
67} 214509 (2003).

\bibitem{thisarticle} M. F. Smith, Johnpierre Paglione, M. B. Walker and Louis Taillefer, Phys. Rev. B {\bf 71}, 014506 (2005).

\end{thebibliography}
\end{document}